\documentclass{nature}

\usepackage{graphicx}
\usepackage{amssymb}
\usepackage{caption}
\usepackage{subcaption}

\usepackage{natbib}

\makeatletter
\let\saved@includegraphics\includegraphics
\AtBeginDocument{\let\includegraphics\saved@includegraphics}
\renewenvironment*{figure}{\@float{figure}}{\end@float}
\makeatother

\title{Atomic-resolution cryo-STEM across continuously variable temperature \\ \\
\emph{``Variable-temperature cryo-STEM''} }

\author{Berit H. Goodge$^{1,2}$, Elisabeth Bianco$^2$, Henny W. Zandbergen$^{3,4}$ \& Lena F. Kourkoutis$^{1,2}$}

\begin{document}

\bibliographystyle{MandM.bst} 

\maketitle

\begin{affiliations}
\item School of Applied and Engineering Physics, Cornell University, Ithaca, NY 14853, USA
\item Kavli Institute at Cornell for Nanoscale Science, Cornell University, Ithaca, NY 14853, USA
\item Kavli Institute of Nanoscience, Delft University of Technology, 2600 GA Delft, The Netherlands
\item HennyZ, 2223 GL Katwijk, The Netherlands
\end{affiliations}

\noindent Correspondence to:\\  \\
Lena F. Kourkoutis \\
School of Applied and Engineering Physics \\
Cornell University \\
142 Sciences Dr. \\
Ithaca, NY 14853\\ 
phone: (607) 255-9121 \\
fax: 607 255-7658\\
email: lena.f.kourkoutis@cornell.edu


\newpage
\begin{abstract}
 Atomic-resolution cryogenic scanning transmission electron microscopy (cryo-STEM) has provided a path to probing the microscopic nature of select low-temperature phases in quantum materials. Expanding cryo-STEM techniques to broadly tunable temperatures will give access to the rich temperature-dependent phase diagrams of these materials. With existing cryo-holders, however, variations in sample temperature significantly disrupt the thermal equilibrium of the system, resulting in large-scale sample drift. 
 The ability to tune temperature without negatively impacting the overall instrument stability is crucial, particularly for high-resolution experiments. 
 
 Here, we test a new side-entry continuously variable temperature dual-tilt cryo-holder which integrates liquid nitrogen cooling with a 6-pin MEMS sample heater to overcome some of these experimental challenges. We measure consistently low drift rates of 0.3-0.4 \AA /s and demonstrate atomic-resolution cryo-STEM imaging across a continuously variable temperature range from $\sim$100 K to well above room temperature. We conduct additional drift stability measurements across several commercial sample stages and discuss implications for further developments of ultra-stable, flexible cryo-stages.  
 
\end{abstract}



\newpage

\section{Introduction}

Following the development of the first electron microscope (\citealp{knoll_elektronenmikroskop_1932}) and the availability of practical commercial instruments in the late 1930s, it was less than two decades before the first experiments to cryogenically cool samples inside a transmission electron microscope (TEM) were reported (\citealp{leisegang_versuch_1954}, \citealp{honjo_low_1956}, \citealp{Fernandez-Moran1960Low-TemperatureIi}). These early demonstrations already recognized the potential of low-temperature electron microscopy to reduce contamination (\citealp{leisegang_versuch_1954}), to image frozen hydrated biological specimens (\citealp{Fernandez-Moran1960Low-TemperatureIi}), and to study low-temperature crystalline phases and phase transitions (\citealp{honjo_low_1956}, \citealp{Fernandez-Moran1960Low-TemperatureIi}). The temperature range was expanded shortly after with the introduction of liquid helium cooled stages (\citealp{Venables1963LiquidMicroscope}, \citealp{Valdre1965AMicroscope}, \citealp{Watanabe1967AMicroscope},  \citealp{Matricardi1967ElectronTribromide}, \citealp{Horl1968}, \citealp{Colliex1968UnLiquide}) which enabled the study of superconductors, solidified gases, and magnetic domains in the 1960s (\citealp{blackman_use_1963}, \citealp{Goringe1964ObservationMicroscopy},  \citealp{Boersch1966ElectronStages}, \citealp{Watanabe1967AMicroscope}, \citealp{Matricardi1967ElectronTribromide}, \citealp{Colliex1968UnLiquide}). Within the materials community, however, the push towards higher resolution over the following decades shifted experimental focus predominantly to more stable (i.e., ambient) working conditions. 

As atomic-resolution imaging and spectroscopy of crystalline materials at room temperature has become almost routine, the interest in cryogenic electron microscopy for the physical sciences has been renewed (\citealp{el_baggari_developments_2019}, \citealp{Zachman2019EmergingMaterials}, \citealp{Minor2019CryogenicScience}). Recent results have demonstrated the potential for cryogenic scanning transmission electron microscopy (STEM) and electron energy loss spectroscopy (EELS) to explore quantum materials phenomena including low-temperature spin states (\citealp{klie_direct_2007}), emergent charge density phases (\citealp{el_baggari_nature_2018}), and interface charge transfer (\citealp{zhao_direct_2018}). Atomic-resolution low-temperature experiments reported to date, however, have been limited to single temperatures set by the choice of cryogen, i.e., liquid nitrogen or helium. Richer exploration of many novel quantum materials will require stable imaging conditions at continuously variable temperatures to tune into phases with narrow stable temperature windows or track phase transitions as they occur.

For this purpose, Zandbergen \emph{et al.} (HennyZ Co.) have developed a continuously variable temperature (CVT) liquid nitrogen specimen holder. The side-entry CVT cryo-holder achieves fine temperature control across the $\sim$100-1000 K range by integrating local heating of the sample via 6-pin MEMS control with liquid nitrogen cooling. Previous measurements with a single-tilt version showed single atom imaging at liquid nitrogen temperature (\citealp{hotz_optimizing_2018}), but dual-tilt capabilities are critical for crystalline materials. Here, we present implementation and performance characterization of a dual-tilt CVT cryo-holder in an aberration-corrected Titan Themis 60-300. We demonstrate sub-\AA \, resolution STEM imaging and low instantaneous drift rates of 0.3-0.4 \AA /s across the temperature range. In addition to the local MEMS heating, this performance is enabled by active heating of the CVT cryo-holder rod which reduces thermal sample drift. We conclude with similar performance analysis of several commercially available room temperature and cryogenic sample stages for comparison and discuss broader trends in this area of instrumentation development.

\subsection{Background}

Side-entry cryo-holders consist of a long hollow metal rod at one end of which the sample is mounted and held in the microscope column. The sample is thermally coupled through the length of the rod to a cryogen cold sink outside the microscope, usually either a dewar or lines of cryogen. Compared to the early dedicated cryo-stages, the introduction of commercial side-entry cryo-holders has made cryo-electron microscopy more widely available (\citealp{henderson_side-entry_1991}). Atomic-resolution STEM with such a setup is, however, complicated by 1) thermal contraction of the macroscopic metal rod which can result in dramatic directional drift of the sample and 2) mechanical coupling of the sample to the the external environment via the cryogen reserve resulting in vibrations from boiling or bubbling within the cryogen. 

Most successful cryo-experiments to date have been performed at cryogen-set temperatures -- usually near $\sim$90 K (liquid nitrogen) or $\sim$10 K (liquid or cold gas helium) -- in order to preserve a fixed thermal equilibrium. In principle, variable temperature control is possible with some commercial cryo-holders through resistive heating of the sample rod (\citealp{tao_role_2011}, \citealp{zhao_direct_2016}, \citealp{berruto_laser-induced_2018}) or by regulation of the cryogen flow through the holder or stage (\citealp{Valdre1965AMicroscope}, \citealp{harada_real-time_1992}, \citealp{qiao_anisotropic_2017}). In practice, the spatial resolution obtained at intermediate temperatures has so far been limited. 

Generally, the most stable imaging conditions are reached after allowing several hours for the temperature throughout the holder, from cryogen to sample, to fully equilibrate. A new equilibrium must be reached whenever the holder is heated to a new temperature, thus requiring additional time to achieve stable conditions after each temperature change. Directional sample drift is also caused by evaporation of the cryogen in the dewar over the course of the experiment. This slow reduction of the cryogen level results in a continuous change to the thermal equilibrium of the system. Attempts to reach intermediate temperatures by heating all or part of the holder rod not only increase the rate at which the cryogen is depleted, but can also lead to bubbling inside the dewar for some target temperatures, further reducing the imaging stability.

Boiling, bubbling, and other disturbances to the cryogen couple directly to the sample through the holder rod (\citealp{henderson_side-entry_1991}), giving rise to a number of imaging artifacts such as ``tearing" of atomic sites in high resolution serial recording techniques such as high-angle annular dark-field (HAADF) STEM (\citealp{klie_variable_2011}). By ensuring proper isolation of the cryogen from the environment and eliminating nucleation sites for bubble formation inside the dewar, random disturbances can be mostly avoided at the base temperature of the holder. Limitations due to the remaining thermal drift were recently overcome by developments in acquisition and processing of fast frame rate STEM data, enabling sub-\AA \, resolution imaging near liquid nitrogen temperature (\citealp{savitzky_image_2018}, \citealp{el_baggari_nature_2018}). The successful expansion of these high resolution techniques to variable temperature \emph{in situ} experiments will be a major step for cryo-STEM applications in physics and materials science.

\section{Materials and Methods}
\subsection{Continuously variable temperature (CVT) cryo-TEM holder}

\begin{figure}
    \centering
    \includegraphics[width=0.75\linewidth]{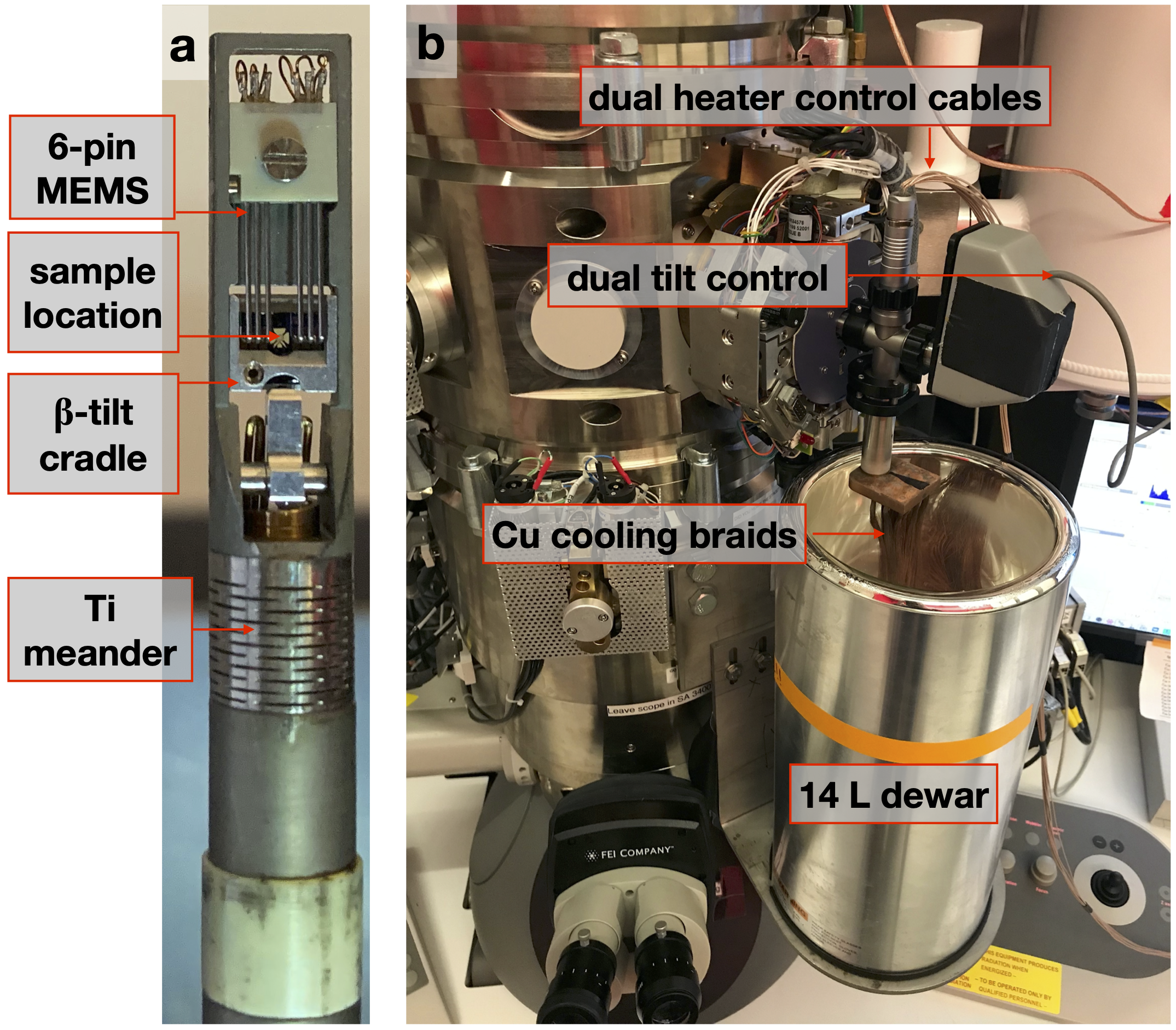}
    \caption{The dual-tilt side-entry continuously variable temperature (CVT) dual-tilt cryo-STEM holder system. (a) A close-up image of the sample rod tip, showing the 6-pin MEMS heater connection and the rigid insulating Ti meander tube. (b) The full set-up of the CVT cryo-STEM holder as implemented in a low-base Titan Themis.}\label{fig:setup}
\end{figure}

The complete HennyZ side-entry CVT cryo-holder system is shown in Figure \ref{fig:setup}. Continuously variable temperature control is achieved through the integration of 6-pin MEMS heating in a side-entry liquid nitrogen cryo-holder. The MEMS chip (Fig. \ref{fig:setup}a) enables local heating directly at the sample without disrupting the thermal equilibrium of the holder. After cooling and equilibrating the entire system, activating the MEMS device heats the sample to the desired temperature above the cryogenic baseline, covering a $\sim$100-1000 K temperature range. All sample temperatures are reported here as measured by the calibrated MEMS chip. 

The CVT cryo-holder also has a ceramic Ti meander tube near the end of the sample rod designed to rigidly connect but thermally isolate the cooled holder tip from the outer cylinder of the rod arm. The cylinder can be heated by a separate temperature controller to match the temperature of the microscope goniometer. This helps maintain thermal equilibrium across the system, thereby further reducing macroscopic drift from expansion or contraction of the rod arm. 

The full set-up includes a dual-axis sample tilt cable connection, external control (not pictured) of both the rod and MEMS heaters, and a large 14 L liquid nitrogen dewar mounted to the side of the microscope column into which the holder's Cu cooling braids are suspended (Fig. \ref{fig:setup}b). 
The large reservoir extends the total imaging time before the cryogen has to be replenished, though the temperature at the sample will slowly increase as the liquid nitrogen level reduces. This change only effectively impacts experiments at the lowest possible temperature, as the MEMS sample heating offsets such variations at intermediate points. 

One drawback of a large dewar is increased sensitivity to external disturbances (\citealp{henderson_side-entry_1991}). Unlike standard side-entry holders, however, the specimen rod is not physically connected to the dewar but only makes contact to the cryogen through the Cu cooling braids. Mounting the dewar directly on the microscope column ensures isolation from other vibrations in the room. The use of fine gauge wire for the cooling braids reduces the strength of mechanical coupling to the liquid nitrogen.

\subsection{Sample drift measurements}
Drift measurements reported in the Results section (see Figs. 4-6) were acquired by measuring shifts via cross correlation between 1-second image frames taken successively over 250 seconds. Sample position plots show the relative motion of the sample with respect to an arbitrary ($x$,$y$) coordinate from the middle frame of the series. Color scales of drift plots indicate frame order from darkest (first) to lightest (last); arrows near the beginning of each series indicate the general drift direction. Instantaneous velocities at each frame were calculated using the leapfrog method (\citealp{savitzky_image_2018}); histograms of their magnitudes accompany each drift plot with solid and dotted vertical lines marking the mean speeds ($\mu$) and standard deviations ($\sigma$), respectively.

\section{Results}

\begin{figure}
    \centering
        \includegraphics[width=\linewidth]{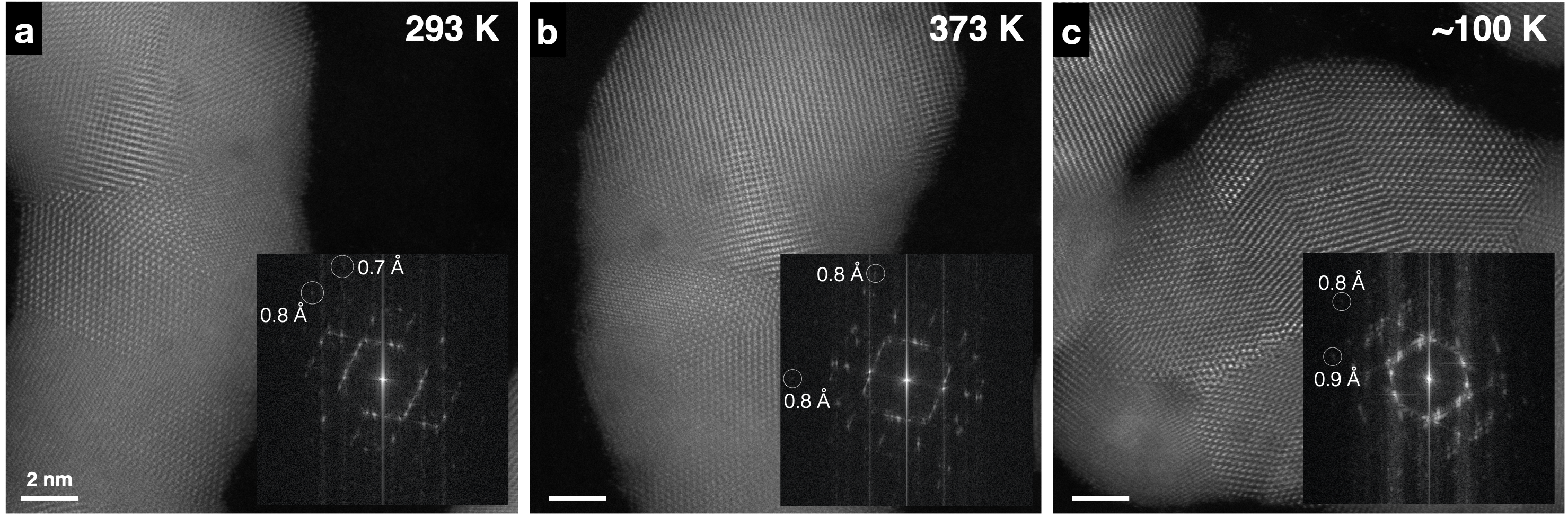}
    \caption{Baseline single-frame HAADF STEM images of polycrystalline Au nanoparticles on C support in the CVT cryo-holder show sub-\AA \,resolution without image registration, drift correction, or other post-processing. (a) Ambient conditions, no cryogen or MEMS heating. (b) No cryogen, MEMS heater at 373 K. (c) Liquid nitrogen cooled to $\sim$100 K, No MEMS heating. Inset fast Fourier transform (FFT) amplitudes show sub-\AA\, information transfer under all conditions. Scans are 2048$\times$2048 px with dwell times of 2 $\mu$s/px for total frame times of $\sim$8.5 s, oriented with slow scan direction horizontal.  }\label{fig:baseline}
\end{figure}

 The stability of the dual-tilt CVT cryo-holder is assessed in STEM mode on an aberration-corrected FEI Titan Themis. Operating at 300 kV, the microscope was tuned to sub-\AA \, resolution at room temperature using a convergence semi-angle of 21 mrad and HAADF collection angles between 68 and 340 mrad. The CVT cryo-holder was loaded into the microscope and allowed to settle for $\sim$1.5 hours at room temperature to minimize the influence of initial drift as the holder settles in the goniometer. Prior to cooling, we performed benchmark tests under ambient condition (Fig. \ref{fig:baseline}a) and with the MEMS chip turned on to heat the sample (Fig. \ref{fig:baseline}b). Cooled to the cryogenic baseline without MEMS heating $\sim$100 K (Fig. \ref{fig:baseline}c), the imaging performance remains nearly unchanged. Small amounts of scan tearing from minor cryogen bubbling is visible in certain regions of the images and as vertical streaking in the fast Fourier transform (FFT) amplitudes. Corresponding inset FFTs show sub-\AA \,information transfer consistently for STEM images at all temperatures. Each scan is 2048$\times$2048 pixels with a 2 $\mu$s/px dwell time for a total frame time of $\sim$8.5 seconds, with all images oriented such that the fast scan direction is horizontal. Imaging resolution can be further improved through fast frame acquisitions coupled with image registration techniques (\citealp{savitzky_image_2018}).

\begin{figure}
    \centering
        \includegraphics[width=\linewidth]{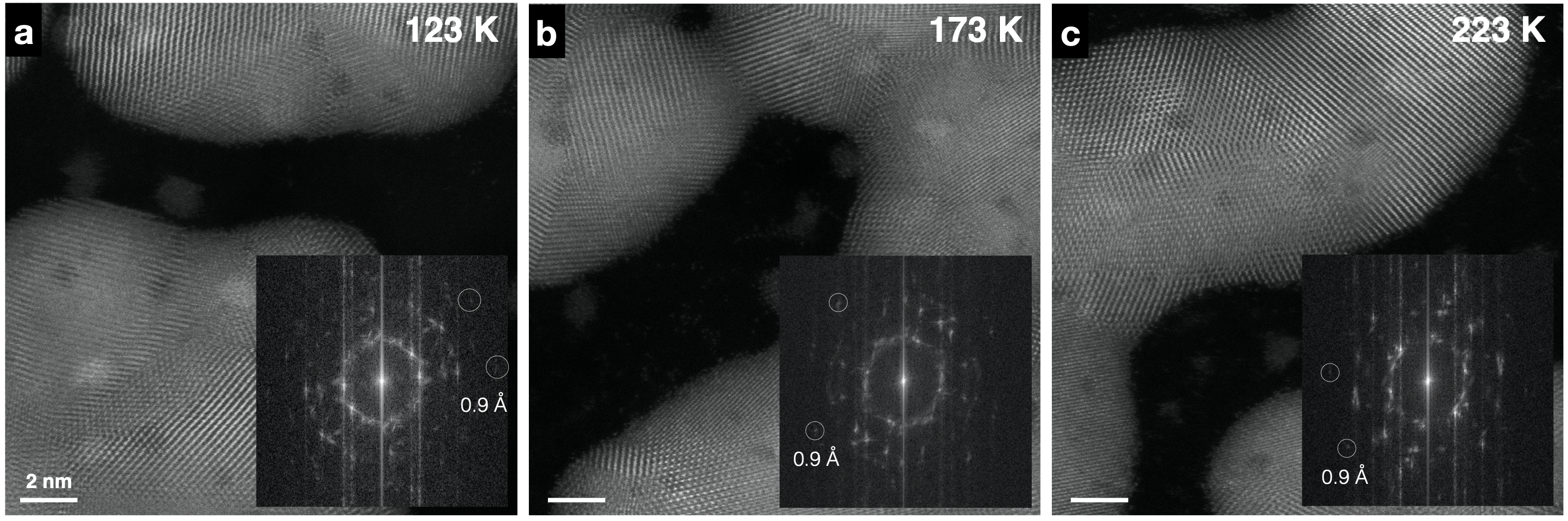}
    \caption{Single-frame HAADF STEM images of polycrystalline Au nanoparticles on C support in the CVT cryo-holder at intermediate temperatures of (a) 123 K, (b) 173 K, and (c) 223 K as measured by the calibrated MEMS chip show sub-\AA \,resolution without image registration, drift correction, or other post-processing. Minimal scan tearing from disturbances in the liquid nitrogen can be observed. Scans are 2048$\times$2048 px with dwell times of 2 $\mu$s/px for total frame times of $\sim$8.5 s, oriented with horizontal slow scan direction.  }\label{fig:int_scan}
\end{figure}
 
Activating the MEMS chip to access continuously variable intermediate sample temperatures results in no observable degradation of imaging stability. Figure \ref{fig:int_scan} shows single-frame HAADF STEM scans of polycrystalline gold nanoparticles deposited on a carbon coated MEMS chip at three intermediate temperatures ($\sim$ 123, 173, and 223 K). These results demonstrate that sub-\AA \, resolution STEM imaging which is today almost standard under ambient conditions can now be achieved at arbitrary cryogenic temperatures down to the base temperature of the CVT cryo-holder.

\begin{figure}
    \centering
        \includegraphics[width=\linewidth]{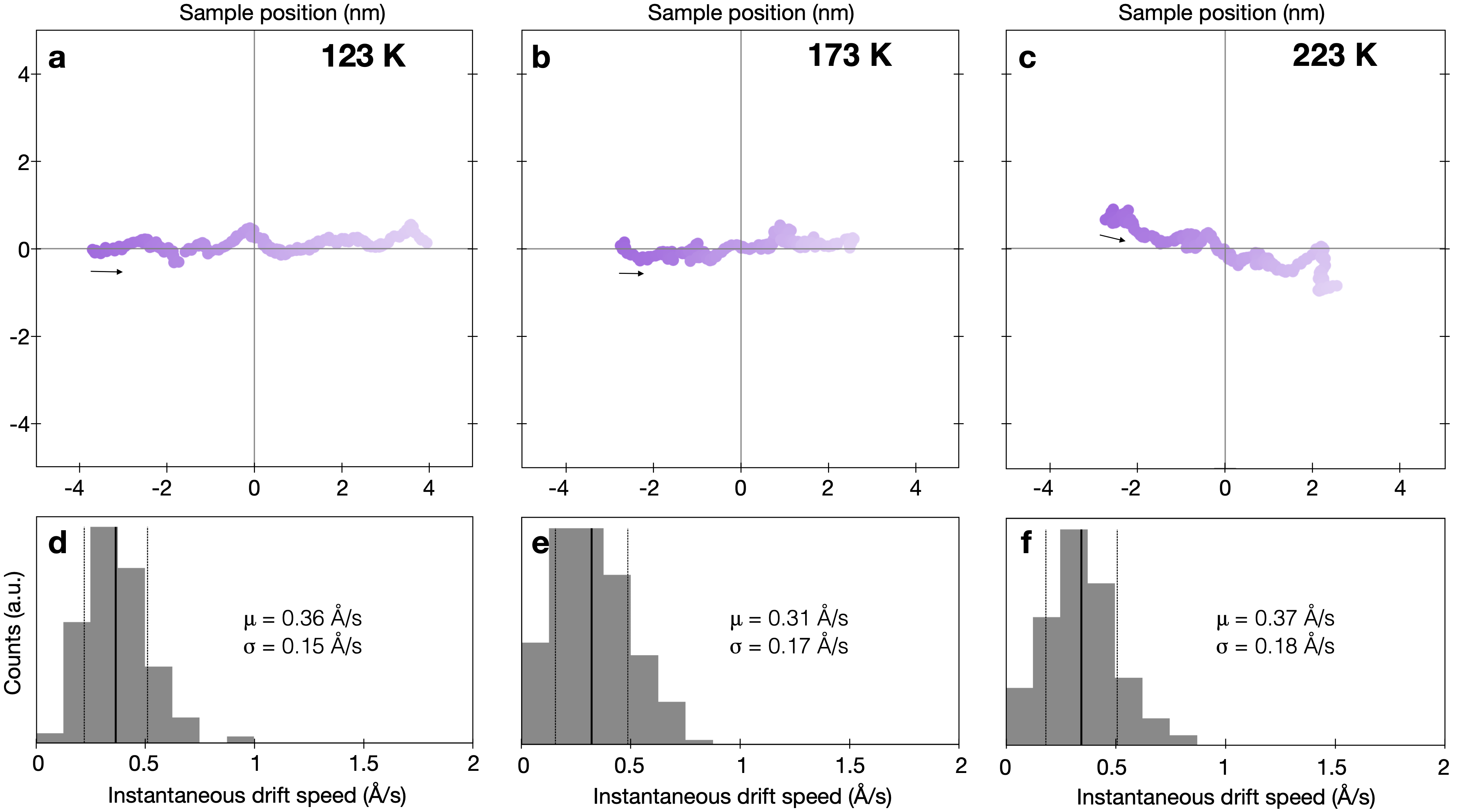}
    \caption{Drift performance of the liquid nitrogen cooled CVT cryo-holder at intermediate temperatures measured by sample positions plotted over 250 s time periods at (a) 123 K, (b) 173 K, and (c) 223 K. (d)-(f) Corresponding instantaneous drift speed histograms with mean speed ($\mu$) and standard deviations ($\sigma$) marked by solid and dotted vertical lines, respectively.}\label{fig:int_drift}
\end{figure}

\begin{figure}
    \centering
        \includegraphics[width=0.66\linewidth]{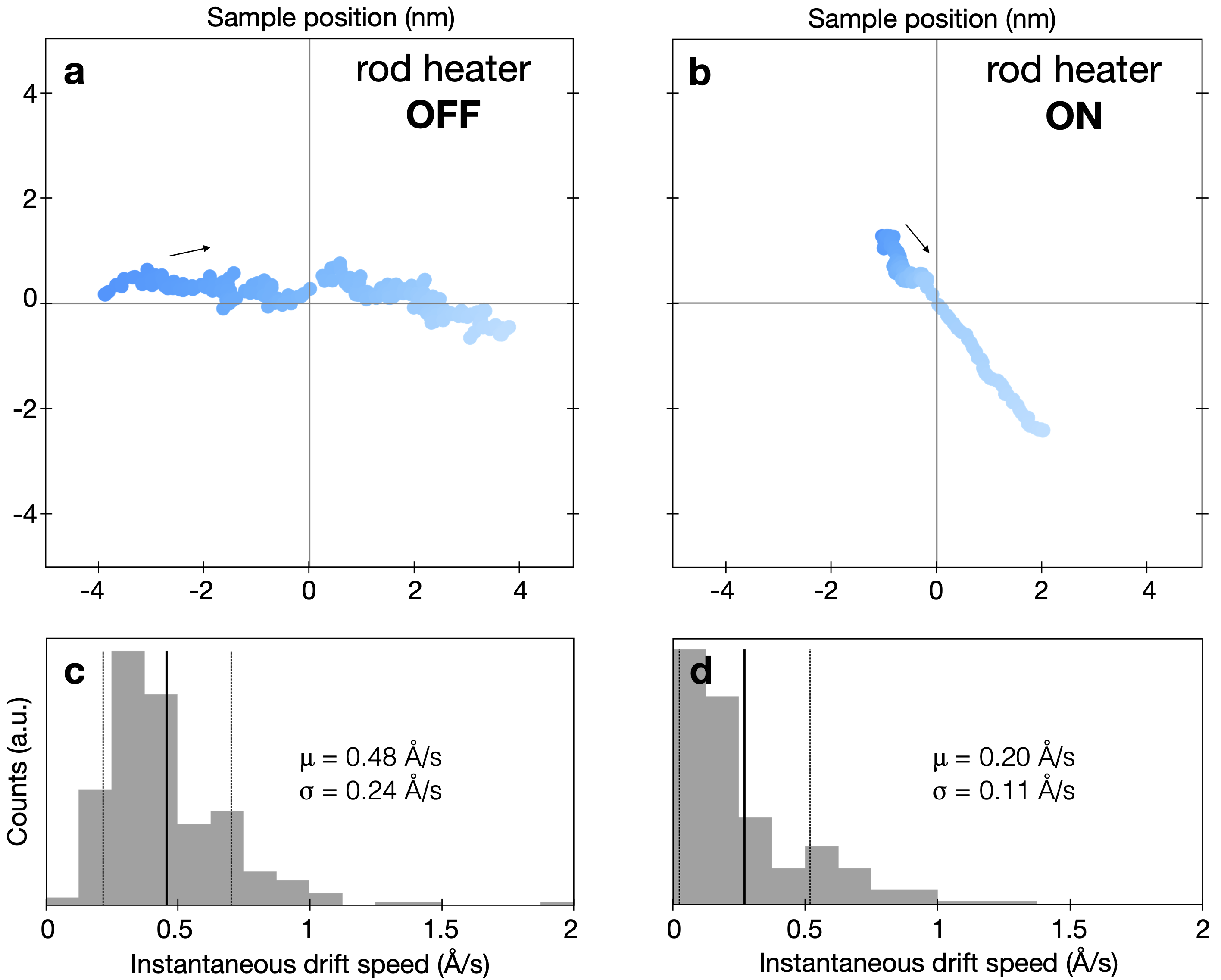}
    \caption{Effect of sample rod heating on drift performance of the liquid nitrogen cooled CVT cryo-holder measured by sample positions plotted over 250 s time periods (a) without and (b) with the rod heater activated. (c, d) Corresponding histograms show a nearly five-fold reduction in instantaneous drift speeds from (a) to (b). Mean speed ($\mu$) and standard deviations ($\sigma$) marked by solid and dotted vertical lines, respectively.}\label{fig:rodheating}
\end{figure}

Local MEMS heating leaves the thermal equilibrium of the holder system effectively unchanged and introduces little to no additional drift when varying the sample temperature. Rapid changes across tens of K require only a few minutes of settle time to reach stable conditions. Figure \ref{fig:int_drift} shows sample drift measurements for the same intermediate temperatures shown in Fig. \ref{fig:int_scan}. Each measurement was taken within 5 minutes after a temperature adjustment of 50 K. The average drift rates, which range between 0.3-0.4 \AA/sec, are only slightly above the cryogenic baseline drift for this holder of 0.2 \AA/sec without MEMS sample heating (see Fig. \ref{fig:rodheating}b,d).

\begin{figure}
    \centering
        \includegraphics[width=0.6\linewidth]{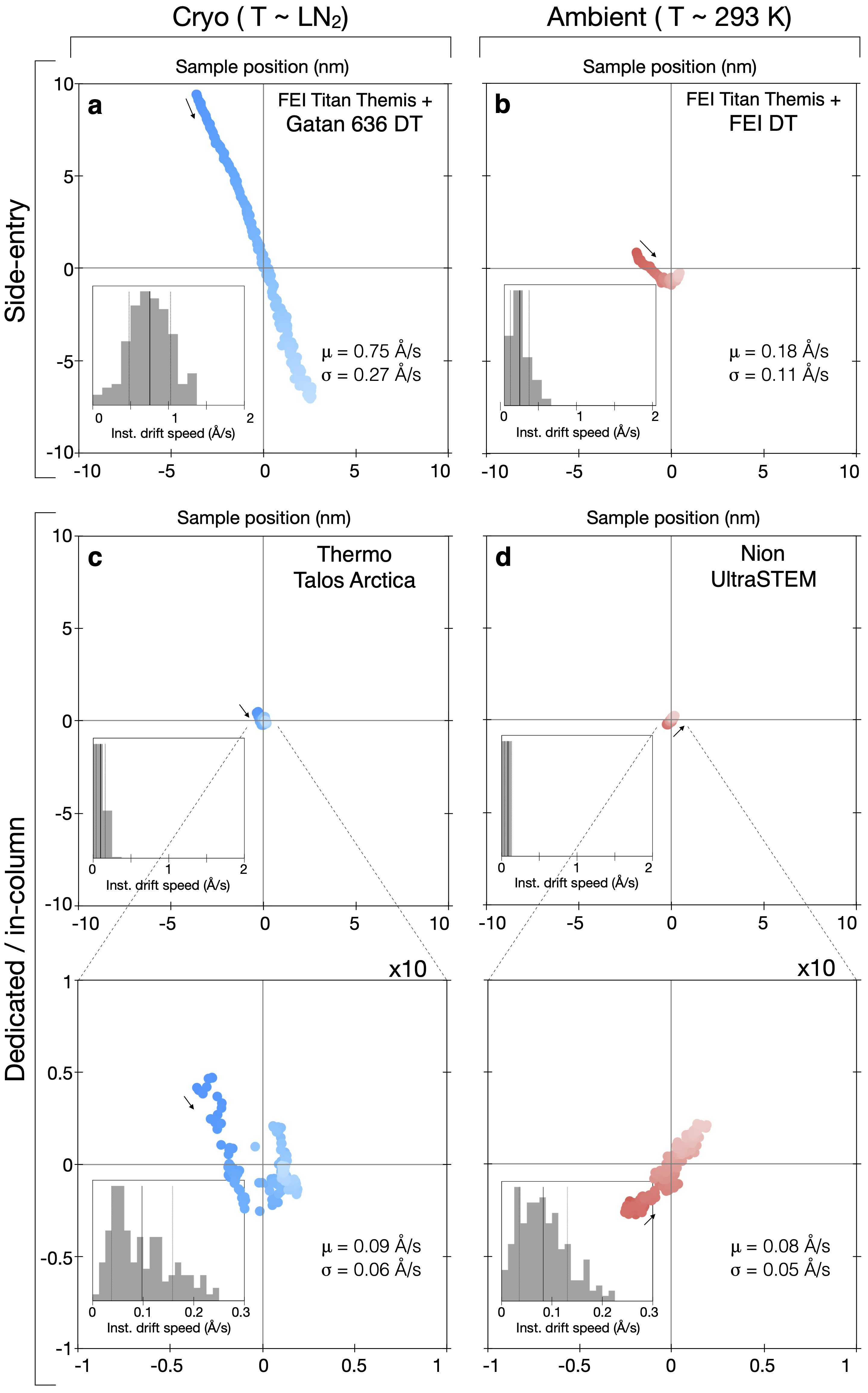}
    \caption{Drift performance comparison of commercial room temperature and cryogenic side-entry holders and dedicated stages. Sample drift plots an corresponding instantaneous drift speed histograms for several commercial sample stages: (a) side-entry dual-tilt Gatan 636 liquid nitrogen cryo holder in a Titan Themis; (b) side-entry dual-tilt FEI room temperature holder in a Titan Themis; (c) dedicated liquid nitrogen cooled cryo-stage in a Thermo Fisher Talos Artica; (d) room temperature Nion UltraSTEM in-column cartridge. 10$\times$ magnifications are included for the (c) Arctica and (d) UltraSTEM. }
\label{fig:dedicated}
\end{figure}

The low drift rates observed across all measured temperatures are enabled in part by the active rod heating and the Ti meander described in Fig. \ref{fig:setup}. Active compensation through the rod heater helps maintain a constant temperature equilibrium of the holder inside the microscope goniometer, reducing the characteristic directional drift by as much as half. Drift measurements in Fig. \ref{fig:rodheating} demonstrate the effect of active heating of the rod to 17.6$^\circ$C, near the measured temperature of the goniometer. The rod heating reduces the average sample drift speed by $\sim$60\%, from 0.48 \AA/sec without rod heating to 0.20 \AA/sec with heating. As a point of comparison, the average sample drift rate of a standard side-entry dual-tilt cryo-holder (Gatan Model 636) in our Titan Themis was measured to be 0.75 \AA/s at its base temperature near liquid nitrogen  (see Fig. \ref{fig:dedicated}a).

The slow depletion of cryogen over the course of a long experiment results in a ``moving target'' equilibrium point as the submersion of the cooling braids slowly decreases. Although a large dewar slows these effects, another way to further compensate this variation is to maintain a constant cryogen level in the dewar so that the cooling braids remain submerged by the same amount over the course of an entire experiment. By mounting an optical monitor inside the nitrogen dewar, it is possible to monitor the level of the cryogen as it boils off and compensate for this depletion by slowly lowering an insulating volume into it, displacing the remaining liquid nitrogen to a constant level. One such system has been tested on a similar CVT cryo-holder in a Philips CM200 with promising results, but no such technique has yet been implemented on the Titan Themis in the experiments described here.

\section{Discussion}

Advances in hardware for cryogenic electron microscopy over the past decades have been primarily driven by needs in the life sciences. This has led to the development of modern dedicated liquid nitrogen cooled stages which support rapid sample exchange, more automated data acquisition, and increase sample throughput (\citealp{Williams2019Cryo-electronScience}). These dedicated cryo-stages, however, lack the dual-axis tilt control which is often indispensable for the physical sciences and currently only available in side-entry cryo-holders. Side-entry holders also offer additional flexibility to more easily incorporate sample manipulation such as electrical biasing or MEMS heating, as in the CVT cryo-holder discussed here. Careful design ensures high stability in both drift and vibrations to enable sub-\AA \, cryo-STEM imaging not only near liquid nitrogen but continuously across a wide temperature range (Figs. \ref{fig:int_scan}, \ref{fig:int_drift}). 

While neither continuous temperature control nor dual-axis tilting is currently possible in dedicated cryo-stages, their superior drift performances compared to side-entry designs is quite striking. Figure \ref{fig:dedicated}a and c show drift measurements from a standard side-entry dual-tilt Gatan 636 cryo-holder in the Titan Themis and a Thermo Talos Artica with dedicated cryo-stage, both near liquid nitrogen temperature. The dedicated cryo-stage offers a nearly ten-fold decrease in sample motion compared to the 0.75 \AA/sec drift of the side-entry cryo-holder. For ambient temperature experiments, the difference in stability between in-column stages and side-entry holders is less severe but still clear. The dual-tilt cartridge of a Nion UltraSTEM in-column stage exhibits very low drift rates of $<$0.1 \AA/sec, less than half the 0.2 \AA/sec drift of a standard FEI dual-tilt side-entry holder in a Titan Themis (Fig. \ref{fig:dedicated}d, b).

Further improvements to the stability of dual-tilt cryo-holders or the development of stable cryo-stages with either dual-tilt or tilt-rotate capabilities will be particularly important for slow-scan techniques such as 4D-STEM and spectroscopic mapping. Some experimental flexibility including sample heating has already been demonstrated with room temperature in-column stages (\citealp{hudak_real-time_2017}, \citealp{sang_situ_2018}), but has as yet not been realized in dedicated cryo-stages. Future developments for high performance specimen holders and stages will ideally seek to combine the benefits of both designs, marrying the ultimate stability of in-column stage designs with the flexibility and variable controls of new \emph{in situ} side-entry holders.

\section{Conclusions}

To date, even successful high-resolution cryo-STEM experiments still make important sacrifices when compared to standard room temperature conditions. Side-entry cryo-holders like those used in (\citealp{klie_direct_2007}, \citealp{el_baggari_nature_2018}, \citealp{zhao_direct_2018}) suffer from instabilities caused by thermal contraction and cryogen boiling, precluding the effective use of longer acquisition techniques such as 4D STEM or spectroscopic mapping. On the other hand, dedicated cryo-stages boast improved stability but at the expense of the dual-axis sample tilt critical for crystalline materials.

The HennyZ CVT cryo-holder system described here demonstrates both drift and internal vibration stability which enables sub-\AA \, STEM imaging across continuously variable temperatures from $\sim$100 - 1000 K. The combination of MEMS sample heating for intermediate temperature access, rod heating to help mitigate thermal drift,  and dual-tilt capabilities offer increased experimental flexibility. These proof-of-concept experiments mark significant progress for the accessibility of variable-temperature cryo-electron microscopy, opening the doors to a new range of high-resolution \emph{in situ} cryo-experiments including real-time observation of phase transitions, temperature cycling, and access to phases which are stable only in narrow temperature windows.
\vspace{1cm}

\section{Acknowledgements}
We thank Katherine A. Spoth for useful discussion and experimental assistance and Ismail El Baggari for providing the data used to collect drift measurements on the Gatan 636 cryo-holder. This work was primarily supported by the National Sciences Foundation, through the PARADIM Materials Innovation Platform (DMR-1539918). We acknowledge additional support by the Department of Defense Air Force Office of Scientific Research (No. FA 9550-16-1-0305) and the Packard Foundation. This work made use of the Cornell Center for Materials Research (CCMR) Shared Facilities, which are supported through the NSF MRSEC Program (No. DMR-1719875). The FEI Titan Themis 300 was acquired through No. NSF-MRI-1429155, with additional support from Cornell University, the Weill Institute, and the Kavli Institute at Cornell. 

\section{Competing Interests} 
B.H.G., E.B., and L.F.K. declare that they have no competing financial interests.
\\
The CVT cryo-holder described here is a product of HennyZ, founded by H.W.Z.

\section{Correspondence} Correspondence and requests for materials
should be addressed to L.F.K. 
\\ (email: lena.f.kourkoutis@cornell.edu)

\end{document}